\documentclass[apj]{emulateapj}
\usepackage{color}
\usepackage[flushleft]{threeparttable}

\begin{document}
\voffset -2cm

\title{A comparative study of host galaxy properties between Fast Radio Bursts and stellar transients}
\author{Ye Li$^{1, 2}$, Bing Zhang$^{3}$}

\affil{
$^1$Kavli Institute for Astronomy and Astrophysics, Peking University, Beijing 100871, China \\
$^2$Purple Mountain Observatory, Chinese Academy of Sciences, Nanjing 210008, China \\
$^3$Department of Physics and Astronomy, University of Nevada, Las Vegas, NV 89154, USA \\
}

\begin{abstract}
Recent arcsecond localizations of Fast Radio Bursts and identifications of their host galaxies confirmed their extragalactic origin. 
While FRB 121102 resides in the bright region of a dwarf star forming galaxy, other FRBs reside in more massive galaxies and are related to older stellar populations.
We compare the host galaxy properties of {nine} FRBs with those of several types of stellar transients: from young to old population, long duration gamma ray bursts (LGRBs), superluminous supernovae (SLSNe), Type Ib/Ic supernovae (SN Ibc), Type II supernovae (SN II), type Ia supernovae (SN Ia), and short duration gamma ray bursts (SGRBs). We find that as a whole sample, the stellar mass and star formation rate of the FRB host galaxies prefer a medium to old population, and are against a young population similar to LGRBs and SLSNe by a null probability 0.02. 
Individually, the FRB 121102 host is consistent with that of young population objects; the FRB 180924 environment is similar to that of SGRBs; and the FRB 190523 environment is similar to those of SN Ia. These results are consistent with the magnetar engine model for FRBs, if both magnetars produced from extreme explosions (GRBs/SLSNe) and from regular channels (e.g. those producing Galactic magnetars) can produce FRBs.
\end{abstract}

\section{Introduction}

Fast Radio Bursts are extragalactic radio transients with durations $0.01-50$ milliseconds (ms) and  dispersion measures (DMs) in excess of the Galactic values \citep{lorimer07,petroff19,cordes19}. Up to now, there are more than 100 FRBs reported 
(FRBCAT\footnote{http://www.frbcat.org/, updated to 2019 December 23} \cite{petroff16}).
While most are one-off bursts, 
at least 20 sources show repeating bursts \citep[e.g.][]{spitler16, scholz16,chime-2ndrepeater,kumar19,chime19-8repeater, chime20-9repeater,luo20}.
More than 50 theoretical models have been proposed
(see \cite{katz16, platts19} for theoretical reviews\footnote{https://frbtheorycat.org/index.php/Main\_Page}). 
Most models invoke neutron stars or other compact objects (e.g. black holes or white dwarfs) as the sources. 

{Thanks to the observations with Karl G. Jansky Very Large Array (VLA), the Deep Synoptic Array ten-antenna prototype (DSA-10),  the European VLBI Network (EVN), and the Commensal Real-time ASKAP Fast Transients Survey (ASKAP/CRAFT), }the arcsecond localization data of nine FRB sources have been published. 
These include the repeating sources FRB 121102 \citep{chatterjee17} and FRB 180916.J0158+65\citep{marcote20} and apparently non-repeating sources FRB 180924 \citep{bannister19}, FRB 181112 \citep{prochaska19}, FRB 190523 \citep{ravi19},
%During the reviewing process of this paper, four more well-localized FRBs, 
{FRB 190102, FRB 190608, FRB 190611 and FRB 190711\citep{bhandari2020, macquart2020}.}
In the field of gamma-ray bursts (GRBs), multi-wavelength properties, especially the host galaxy properties, have played an important role in identifying two physically distinct classes of the sources, i.e. long GRBs (LGRBs) due to core collapse of massive stars and short GRBs (SGRBs) due to binary neutron star mergers \citep[e.g.][]{2006Natur.441..463F,fong10,berger14,blanchard16,2016ApJS..227....7L}.
For FRBs, the host galaxy properties and the location of the FRB source within the galaxies also carry the clue to diagnose their possible origin(s).

The properties of the FRB host galaxies so far indicate a perplexing picture. 
FRB 121102, the first repeater and the first FRB localized with an arcsecond precision, 
resides in the brightest region of a star forming dwarf galaxy
\citep{tendulkar17, bassa17, chatterjee17,kokubo17}, whose properties are quite similar to the host galaxies of young stellar population transients, e.g., LGRBs and Superluminous Supernovae (SLSNe). This observation motivated the suggestion that young magnetars produced from these extreme explosions are the sources of repeating FRBs
\citep[e.g.][]{murase16,metzger17, beloborodov17}, and it was predicted that the majority of FRBs should reside in the similar environments \citep[e.g.][]{nicholl17}. 
However, later {precise} localizations of other FRBs suggest otherwise. For example,
FRB 180924, FRB 190523 and FRB 180916.J0158+65 are located in massive galaxies. Furthermore, FRB 180924 resides far away from the center of its host. These properties are similar to those of the host galaxies of old population transients, such as SGRBs \citep{bannister19,ravi19}.
This may point towards an origin of FRBs related to compact binary coalescences \citep[e.g.][]{totani13,margalit19,zhang20,wang20}.

It is possible that FRBs are not related to the extremely young or extremely old stellar populations. If this is the case, then neither LGRBs/SLSNe nor SGRBs are good population proxies of FRBs. It is also possible that the observed FRBs may include sub-classes with diverse origins, as most other astrophysical phenomena do. Indeed, the FRB host properties seem to be diverse given the limited information available \citep{liye19}. In order to make an assessment to the origin of FRBs based on their host galaxy data, it is essential to collect the statistical properties of the host galaxies of different types of stellar transients and cross compare the FRB host properties with them. 

In this paper, we carry out such a task. Besides LGRBs/SLSNe and SGRBs that represent the youngest and oldest stellar populations, we also perform a statistical analysis of the host galaxies of intermediate stellar transients. 
From young to old, they are LGRBs, SLSNe, Type Ib/Ic supernovae (SN Ibc), Type II supernovae (SN II), type Ia supernovae (SN Ia), and SGRBs. We compare the host properties between FRBs and these transients trying to address the following questions: Which host galaxy type the FRB hosts are more analogous to as a whole or individually? Could there be diverse origins of FRBs?
We construct the paper as follows.
The host galaxy samples of different types of transients are presented in Section 2. 
The host galaxy properties of FRBs are compared with those of different types of stellar transients as a whole in Section 3 and individually in Section 4.
The implications of our results are discussed in Section 5. 
The cosmological parameters $H_0=67.8$ km s$^{-1}$ Mpc$^{-1}$, $\Omega_{\rm m}=0.308$, and  $\Omega_{\rm m}=0.692$
are adopted \citep{planck16}.

\section{Parameters and Samples}\label{chp_sample}

We discuss the parameters considered in this paper in Section \ref{parameters}
and present the samples of FRBs and various transients in Section \ref{objects}.

\subsection{Parameters} \label{parameters}

Our goal is to compare the host galaxy properties of FRBs with those of other transients. The properties of the host galaxies can be documented in a set of parameters, both the global properties in the galactic scale and the local properties in the sub-galactic scale. We discuss these parameters in turn. 

\subsubsection{Galactic-scale parameters: log $M_*$, SFR, metallicity, $R_{50}$}

The most important global properties of the host galaxies are stellar mass $M_*$,
star formation rate SFR, specific star formation rate sSFR (SFR/$M_*$), metallicity 12+log(O/H), 
as well as the half light radius $R_{50}$ of the host galaxies.
The stellar mass $M_*$ of a galaxy is usually estimated by broadband spectral energy distribution (SED)
fitting to the stellar population synthesis \citep{SED_LePhare, 2003MNRAS.341...33K, SED_CIGALE}.
Star formation rate is estimated by emission lines such as H$\alpha$, or ultra violet (UV) luminosity (see \cite{2012ARA&A..50..531K} for a review).
sSFR is estimated as SFR/$M_*$ when SFR and $M_*$ are both available.
The transients related with younger populations (LGRBs and SLSNe) usually reside in the 
galaxies with smaller stellar masses and more intense SFR and sSFR (and sometimes less metallicity) than those related to old populations
\citep{2008ApJ...687.1201K, schulze2018}.
The metallicity of a galaxy is usually estimated with the emission 
\citep{2002ApJS..142...35K, 2008ApJ...681.1183K, 2004ApJ...617..240K, 2004MNRAS.348L..59P, 2016Ap&SS.361...61D}, 
or absorption line ratios \citep{2011piim.book.....D}.
The emission line method gives metallicity in the form of $12+\rm log(O/H)$, with the solar metallicity being $12+\rm log(O/H)_{\odot}=8.69$ \citep{2009ARA&A..47..481A}.
The absorption line method, on the other hand, gives metallicity in the form of 
${\rm [X/H]} = {\rm log}(N_{\rm X}/N_{\rm H}) - {\rm log}(N_{\rm X}/N_{\rm H})_{\odot}$, where $N_{\rm X}$ indicates the column density of element X. 
To be consistent, we convert $12+\rm log(O/H)$ to $\rm [X/H]$ in this paper.
For those estimated with emission lines, we choose to use those estimated based on \cite{2016Ap&SS.361...61D} when available, 
to be consistent with that estimated for FRB 180916.J0158+65.

The half light radius $R_{50}$ is the radius that encloses 50\% of the total light of the galaxy.
It is usually estimated by fitting the surface brightness of a galaxy with
the S{\'e}rsic profile $$\Sigma(r)=\Sigma_{\rm e} {\rm exp}\{{-k_{\rm n}[(r/r_{\rm e})^{1/n}-1]}\},$$
where the effective radius $r_{\rm e}$ represents $R_{50}$. 
Another way is to fit the brightness profiles with ellipses centered around the galaxy and identify the one whose enclosed flux is half of the total flux. $R_{50}$ is defined as the semi-major axis of the ellipse.
In general, $R_{50}$ scales with stellar mass $M_*$. For the same stellar mass,
a star-forming galaxy usually has a larger $R_{50}$ than a passive galaxy.

\subsubsection{Sub-galactic parameters: $R_{\rm off}$, $r_{\rm off}$, $F_{\rm light}$}

The same galaxy may host different types of transients. 
Thus, local properties at the sub-galactic level can provide more precise diagnostics to the environment of a certain transient. One important property is the offset of the transient from the center of the host galaxy. It can be measured in the physical units (kpc) as $R_{\rm off}$, or normalized to the characteristic radius of the host $r_{\rm off}=R_{\rm off}/R_{50}$. The larger the $R_{\rm off}$, the farther away the transient is from the center of the host, and the fainter and more quiescent the local environment is.
However, $R_{\rm off}$ is misleading for irregular galaxies since the center of the galaxy is hard to define and usually does not mark the region with most intense star formation. For these cases, $F_{\rm light}$ is a more efficiency parameter. It is defined as the total light emitted in the region fainter than the transient position, within the host. By definition, a transient within the brightest region would have $F_{\rm light} \sim 1$, and that within the faintest region would have a $F_{\rm light} \sim 0$ \citep{2006Natur.441..463F, 2015MNRAS.448..732A, 2015PASA...32...19A}.

For nearby transient, surface brightness $\Sigma$, local color, local star formation rate density $\Sigma_{\rm SFR}$ would give more precise information. However, these parameters are not available for most objects at larger distances. 
Since the redshifts of the localized FRBs are in the range of $0-1$, a valid local star formation rate density $\Sigma_{\rm SFR}$ is hard to obtain for most of them. We therefore use galaxy-scale properties $M^*$, SFR, sSFR, $\rm [X/H]$, and $R_{50}$, and sub-galactic scale properties $R_{\rm off}$, $r_{\rm off}=R_{\rm off}/R_{50}$, and $F_{\rm light}$ in this study.

\subsection{Samples of FRBs and Stellar Transients \label{objects}}

Different types of transients show somewhat different properties in both global galactic and sub-galactic features \citep{2006Natur.441..463F, 2008ApJ...687.1201K, 2012ApJ...759..107K, 2012MNRAS.424.1372A, 2016ApJS..227....7L}. Comparing the properties of FRBs with those of other transients can shed light on the origin of FRBs. In the following, we discuss the samples of FRBs and other transients used in our study. The sample size for each parameter for each type of stellar transient is presented in Table \ref{tbsample}. 

\subsubsection{Fast Radio Bursts}
\begin{table*}[!htb] \scriptsize
\begin{center}
\caption{Host galaxy properties of FRBs with host galaxies identified.}
\label{tbfrb}
\begin{tabular}{l|ccccccccccc}
\hline
 & instrument & $z$ & log SFR  & log sSFR  & log $M_*$  & $\rm [X/H]$ & $R_{50}$  & offset  & offset  & $F_{\rm light}$ & reference \\\hline
 & &  & M$_{\odot}$ yr$^{-1}$ & Gyr$^{-1}$ & $M_{\odot}$ &  & kpc & kpc & $R_{50}$ &  & \\\hline
FRB121102 & Arecibo & 0.19273 & -0.40 & 0.86 & 7.7 & -0.59 & 1.4 & 0.82 & 0.60 & 1.0 & 1 \\
FRB180916.J0158+65 & CHIME/FRB & 0.0337 & -1.0 & -2.0 & 10.0 & 0.13 & 3.3 & 4.7 & 1.5 & ... & 5,7 \\
FRB180924 & ASKAP & 0.3214 & $<$0.30 & $<$-2.0 & 10.3 & ... & 2.8 & 3.8 & 1.4 & 0.08 & 3 \\
FRB181112 & ASKAP & 0.4755 & -0.22 & -0.62 & 9.4 & ... & 3.9 & 3.1$^{+15.7}_{-3.1}$ & 0.79 & ... & 4,6 \\
FRB190102 & ASKAP & 0.2913 & 0.18 & -0.33 & 9.5 & -0.25 & 5.3 & 1.5$^{+3.4}_{-1.5}$ & 0.28 & ... & 6 \\
FRB190523 & DSA-10 & 0.66 & $<$0.11 & $<$-2.0 & 11.1$\pm$0.1 & -0.52 & ... & 26.5$^{+15.5}_{-14.9}$ & ... & ... & 2 \\
FRB190608 & ASKAP & 0.11778 & 0.079 & -1.3 & 10.4 & -0.34 & ... & 6.8$\pm$1.3 & ... & ... & 6 \\
FRB190611 & ASKAP & 0.378 & ... & ... & ... & ... & ... & 17.2$\pm$4.9 & ... & ... & 8 \\
FRB190711 & ASKAP & 0.522 & ... & ... & ... & ... & ... & 1.5$^{+3.6}_{-1.5}$ & ... & ... & 8 \\
\hline \end{tabular}
\end{center}
\footnotesize
Reference:
(1) \cite{tendulkar17};
(2) \cite{ravi19};
(3) \cite{bannister19};
(4) \cite{prochaska19};
(5) \cite{2019arXiv191202905A};
(6) \cite{bhandari2020};
(7) \cite{marcote20};
(8) \cite{macquart2020}.
\end{table*}

We summarize the host galaxy properties of the FRBs studied in this paper in Table \ref{tbfrb}.
For FRB 180916.J0158+65, the global SFR is scaled from that in the FRB position as
${\rm SFR}=0.016*6.57/1.002=0.1 M_{\odot}\ {\rm yr}^{-1}$.
The metallicity of FRB 180916.J0158+65 is estimated based on \cite{2016Ap&SS.361...61D}.
We then choose to use the metallicity values based on the same reference for FRB 121102 as well as other transients when available.
The half light radius $R_{50}$ of FRB 180916.J0158+65 is not available from the paper.
We estimate $R_{50}$ with the Petrosian half-light radius $R_{50,\rm petro}=4.66"$ and the 90\%-radius as $R_{90,\rm petro}=8.53"$ 
from SDSS catalog, using the formula
$R_{50}=R_{50, \rm petro}/(1-8\times 10^{-6}(R_{90, \rm petro}/R_{50, \rm petro})^{8.47})=4.7"$
\citep{2005AJ....130.1535G}. This gives a physical distance of 3.3 kpc.
We estimate the offset of FRB 190523 with the coordinates (J2000) of the FRB, right ascension (RA) 13:48:15.6(2), declination (DEC) $+$72:28:11(2), and the host coordinates from the PanSTARRS, stack RA 13:48:15.426 (207.06427 degree), DEC $+$72:28:14.6 (72.47072 degree). The offset is estimated to be $3.7^{+2.2}_{-2.1}$ arcsec, corresponding to $26.5^{+15.5}_{-14.9}$ kpc for its redshift. {Note that the astrometric registration between FRB 190523 and the Pan-STARRS image is not available here, 
which may result in an additional $\sim (0.3"-0.5")$ uncertainty.}

%During the reviewing process of this paper, four more FRB hosts are published \citep{bhandari2020, macquart2020}. We add these four cases in our updated sample. 
{The metallicities from \cite{bhandari2020} are provided as $Z$. They} are converted to [X/H] by [X/H]=log$_{10}$($Z/Z_{\odot}$), where $Z_{\odot}=0.0196$ is the solar metallicity.

In summary, {seven} well-localized FRBs have stellar mass, SFR, and sSFR available.
All of the nine FRBs have offset information available. Note that while FRB 121102, FRB 180916.J0158+65, FRB 180924, FRB 190608, and FRB 190611 have well defined offsets, the offsets of FRB 181112, FRB 190102 and FRB 190711 are consistent with being zero. Although the localization of FRB 190523 is relatively poor, {a relatively large offset is favored.}

\subsubsection{LGRBs and SGRBs samples}

\cite{2016ApJS..227....7L} compiles 407 GRBs with redshifts or host galaxy properties from the literature (e.g. \citealt{2006Natur.441..463F,fong10,berger14,blanchard16} and references therein), with all the estimation methods labelled.
%{Their host galaxy properties include all galactic and sub-galactic properties here. Their stellar masses ($\log M_*$) are estimated with SED fitting and IR luminosities; SFR is estimated with emission lines, as well as UV and IR luminosities; Metallicity is mainly estimated with both emission lines and afterglow absorption lines in some cases; Half light radius $R_{50}$ is mainly  estimated with the S{\'e}rsic profile, with some converted from $R_{80}$;  Offset $R_{\rm off}$ and cumulative light fraction $F_{\rm light}$ are  produced by the standard methods. The redshifts of the bursts are mainly determined by emission or absorption lines with photometric redshifts adopted for some GRBs. 
{The stellar mass ($M_*$) values of \cite{2016ApJS..227....7L} are mainly obtained from SED-fitting by \cite{2009ApJ...691..182S} and \cite{2010ApJ...725.1202L}, but some are estimated from the K band or infrared (IR) magnitudes. Due to the large uncertainty of the latter method, only those estimated with SED fitting are included in this study.
The SFR values collected in \cite{2016ApJS..227....7L} are mainly estimated with emission lines from \cite{2009ApJ...691..182S}, \cite{2015A&A...581A.125K} and \cite{2009ApJ...690..231B}, with some estimated using UV fluxes. The emission lines usually trace recent star formation ($<10$ Myr), related to LGRBs, while the UV fluxes trace 10-200 Myr star formation, related to core collapse SNe (See \cite{2012ARA&A..50..531K} for a review). Thus, we use the SFR values estimated using both methods. The metallicity values collected in \cite{2016ApJS..227....7L} are estimated with both emission line ratios \citep{2004MNRAS.348L..59P, 2004ApJ...617..240K, 2008ApJ...681.1183K, 2009ApJ...691..182S, 2015A&A...581A.125K} and absorption line ratios \citep{2011piim.book.....D, 2015ApJ...804...51C}. If the metallicity [X/H] is estimated with $R_{23}=\rm ([OII]\lambda3727+[OIII]\lambda4959, 5007)/H\beta$, the results are double-valued from these two references: \cite{2008ApJ...681.1183K, 2009ApJ...691..182S}. In this case, we only select the larger value of the two following  \cite{2004ApJ...617..240K} and \cite{2009ApJ...690..231B}.
The host galaxy half light radius $R_{50}$ and the offset of the GRB from the center of the host are  mainly derived from HST images \citep{2002AJ....123.1111B, 2007ApJ...657..367W, 2010ApJ...708....9F, 2013ApJ...776...18F, 2016ApJ...817..144B}.
Usually galaxies are inclined. The offsets are sometimes corrected for the inclination. However, the statistical results are not influenced significantly by the inclinations of the objects \citep{2018A&A...617A.105J}. 
All the FRB and GRB offsets are projected offsets. To be consistent, we also use projected offsets in the SN section. 
See \cite{2016ApJS..227....7L} for more details. }

Here we only use the well measured parameters, with upper and lower limits excluded.
%For redshifts, only the spectroscopically identified ones are selected, with photometric redshifts excluded. 
%For the host galaxy stellar mass $M_*$, only those estimated with SED fitting are included, with those estimated using infrared (IR) luminosity (usually with a large uncertainty) excluded. 
Following \cite{liye2020}, here we do not include the four GRBs whose physical categories are subject to debate, i.e. GRB 060505, GRB 060614, GRB 090426, and GRB 060121. 
{The numbers of LGRBs and SGRBs with host galaxy properties are 263 and 31, respectively. To make the GRB sample more consistent with that of FRBs, we only use LGRBs and SGRBs with redshifts $z<1$ for comparison. This results in smaller sample sizes, i.e. 72 and 22 for LGRBs and SGRBs, repectively.}

\begin{table}[!bht] \scriptsize
\begin{center}
\caption{Sample size for each parameter of LGRB, SLSNe, SN Ibc, SN II, SN Ia and SGRB.}
\label{tbsample}
\begin{tabular}{l|cccccc}
\hline
type & LGRB & SLSNe & SNIbc & SNII & SNIa & SGRB \\
total & 371 & 195 & 1300 & 6137 & 13086 & 32 \\
log $z$ & 349 & 190 & 1218 & 4099 & 12290 & 24 \\
log SFR (M$_{\odot}$ yr$^{-1}$) & 200 & 93 & 309 & 1122 & 2337 & 20 \\
log sSFR (Gyr$^{-1}$) & 92 & 93 & 302 & 1085 & 2317 & 19 \\
log $M_*$ ($M_{\odot}$) & 98 & 93 & 376 & 1257 & 2658 & 22 \\
$\rm [X/H]$ & 131 & 28 & 340 & 1290 & 2650 & 9 \\
log $R_{50}$ (kpc) & 126 & 25 & 574 & 1868 & 4837 & 22 \\
log offset (kpc) & 134 & 35 & 767 & 2403 & 4915 & 26 \\
log offset ($R_{50}$) & 115 & 20 & 511 & 3041 & 3293 & 22 \\
$F_{\rm light}$ & 97 & 16 & 101 & 190 & 163 & 18 \\
\hline
any host parameter & 263 & 107 & 900 & 4251 & 7221 & 31\\
sSFR \& $M_*$ & 92 & 93 & 300 & 1078 & 2317 & 19 \\
all parameter & 26 & 2 & 52 & 95 & 70 & 6\\\hline
$z<1$ \\\hline
log $z$ & 91 & 167 & 1217 & 4099 & 12227 & 23 \\
log SFR (M$_{\odot}$ yr$^{-1}$) & 64 & 81 & 306 & 1101 & 2304 & 19 \\
log sSFR (Gyr$^{-1}$) & 49 & 81 & 300 & 1065 & 2286 & 18 \\
log $M_*$ ($M_{\odot}$) & 53 & 81 & 373 & 1233 & 2597 & 19 \\
$\rm [X/H]$ & 42 & 28 & 337 & 1265 & 2589 & 9 \\
log $R_{50}$ (kpc) & 48 & 19 & 564 & 1868 & 4828 & 16 \\
log offset (kpc) & 51 & 29 & 758 & 2403 & 4914 & 18 \\
log offset ($R_{50}$) & 45 & 14 & 497 & 1623 & 3131 & 16 \\
$F_{\rm light}$ & 32 & 10 & 101 & 190 & 163 & 13 \\
\hline
any host parameter & 72 & 95 & 874 & 2829 & 7003 & 22\\
sSFR \& $M_*$ & 49 & 81 & 298 & 1058 & 2286 & 18 \\
all parameter & 12 & 2 & 52 & 95 & 70 & 6\\
\hline
\end{tabular}
\end{center}
\end{table}

\subsubsection{SLSNe, SN Ibc, SN II, SN Ia samples}
For SNe,  we use the data from the Open Supernovae Catalog (OSC)\footnote{https:\/\/sne.space} as the starting point to build our samples. 
The OSC includes the coordinates (RA and DEC) of the SNe and the names and coordinates of their hosts.
Some host galaxies in the OSC do not have coordinates labelled. For these,
we search for their names in SIMBAD\footnote{http://simbad.u-strasbg.fr/simbad/sim-fid} to collect their coordinates.
To be consistent, all available host galaxy names and coordinates are calibrated to SIMBAD ID and coordinates. 
Some host galaxies are not available in SIMBAD. They are calibrated to NED\footnote{https://ned.ipac.caltech.edu/forms/gmd.html} instead. 
In order to reduce the mis-identification of the host galaxies, we exclude those SNe with  host galaxy distances larger than 1 degree.
{We then collect the host galaxy information from the papers exploring SNe and galaxy catalogs.}

\paragraph{Properties from SN papers}

We first supplement the host galaxy properties from the papers exploring the SN host galaxy properties.
We match their SN names with the OSC names or aliases during this process.

For the stellar mass, we use the values estimated with SED fitting only.
For SFR, we prefer those estimated by emission lines, especially H$\alpha$. If it is not available, we use the value estimated using the far ultraviolet (FUV) method. When no value from the above two methods is available, we use the SFR value derived from the SED fitting.
\cite{2012ApJ...759..107K} estimated the stellar mass using SED fitting,
sSFR using fiber spectra, and metallicity using emission lines and the PP04 method for different types of transients.
\cite{taggart2019} and \cite{schulze2018} estimated the stellar mass and SFR of core-collapse supernovae and SLSNe using SED fitting.
For SLSNe only, \cite{2016ApJ...830...13P} estimates  log $M_*$ and SFR using SED fitting and estimated the SFR with emission lines when spectra are available.
For SNe Ia, more than 600 log $M_*$, SFR and metallicities are produced using SED fitting by \cite{2018ApJ...854...24K}.

{For metallicity, we prefer those values estimated using the emission line method, especially those based on \cite{2004MNRAS.348L..59P} (PP04) and \cite{2016Ap&SS.361...61D} (D16).
To be consistent with FRB180916.J0158+65, whose host metallicity was estimated with D16, we adopt those from D16 whenever available. Otherwise, the values from PP04 are adopted.
\cite{2019arXiv190513197G} estimated the SN host metallicities 
provided the largest SN sample estimated with D16.
\cite{2012ApJ...759..107K}, \cite{2019MNRAS.490.4515S}, and \cite{2016A&A...589A.110A} estimated metallicities with the PP04 method for various transients.
For SN Ia located in passive galaxies, \cite{2016ApJS..223....7K} and \cite{2019arXiv191204903K} estimated the host metallicities using absorption line ratios.
We use those [M/H] values based on the Yonsei evolutionary population synthesis (EPS) models \citep{2013ApJS..204....3C}.
}

We prefer the half-light radius $R_{50}$ estimated using 2D S{\'e}sic fitting.
\cite{2015ApJ...804...90L} and \cite{2018A&A...617A.105J} provided $R_{50}$ in $r$ band for SLSNe and SNe Ic-BL, respectively. Other $R_{50}$ values of SN hosts are obtained from galaxy catalogs.

The offsets $R_{\rm off}$ between SNe and their hosts are usually available from the OSC.
However, the offsets are sometimes not trustful, because the RA of the host galaxies in the Asiago SN catalog is accurate to seconds, with an uncertainty of 15 arcseconds. These  are not suitable for offset calculations. 
We thus extract the host galaxy coordinates from the SDSS-II catalog \citep{sako2018},
the ASAS-SN catalog\footnote{http://www.astronomy.ohio-state.edu/~assassin/sn\_list.html}, and the bright SN catalog\footnote{http://www.rochesterastronomy.org/snimages/snredshiftall.html} by matching the OSC SN names with the names in other catalogs.
%Usually galaxies are inclined. The offsets are sometimes corrected for the inclination. However, the statistical results are not influenced significantly by the inclinations of the objects \citep{2018A&A...617A.105J}. 
In our analysis, the offsets for the FRB sample as well as the OSC and GRB samples are not corrected for inclination. We therefore use the projected offsets whenever available. 
We also update the offset values from the detailed papers exploring SN offsets. In particular,
\cite{2016A&A...589A.110A} and \cite{2018A&A...617A.105J} provided the projected $R_{\rm off}$.
\cite{2012ApJ...759..107K} gave the de-projected $r_{\rm off}$.

The $F_{\rm light}$ values are  compiled from the papers exploring SN properties.
\cite{2012MNRAS.424.1372A} calculated $F_{\rm light}$ in both H$\alpha$ and nUV bands for various SNe transients.
H$\alpha$ traces on-going star formation ($0-16$ Myr), which is more relevant to LGRBs,
and nUV traces recent star formation ($16-100$ Myr)\citep{2009ApJ...691..115G}, which is more relevant to core collapse supernovae.
However, the $F_{\rm light}$ values for H$\alpha$ are usually not available for relatively high redshifts.
We thus employ the $F_{\rm light}$ for nUV whenever available.
\cite{2015MNRAS.448..732A} and \cite{2015ApJ...804...90L} provided the $F_{\rm light}$ values from UV for SNe Ia and SLSNe, respectively.  
\cite{2008ApJ...687.1201K} estimated the $F_{\rm light}$ from the $g$ band.

\paragraph{Properties from galaxy catalogs}

We then supplement the host galaxy properties from the catalogs of galaxies.
{During this process we match galaxies with SN hosts both by coordinates with a 3" precision and names.}

For the galaxies within the coverage of SDSS, we use the parameters derived from the SDSS spectrum and broadband photometrics.
The MPA-JHP group provided the stellar mass $M_*$, SFR, and metallicity of SDSS DR8 galaxies by taking both spectrum and
photometrics into account \citep{2003MNRAS.341...33K}.
We use their results when available.
However, they did not provide the results for galaxies later than DR8.
The Flexible Stellar Popullation Synthesis (FSPS) used the SPS method to estimate the galaxy properties for both DR8 and DR12 galaxies
\citep{2009ApJ...699..486C}. We use their results for those not available in the MPA-JHU catalog. 
\cite{2013AJ....145..101K}\footnote{https://www.sao.ru/lv/lvgdb/introduction.php}
collected the SFR values estimated from H$\alpha$ and FUV of galaxies in the Local Volume.
Up to 2020 Jan 1st, it includes 1212 galaxies.
We calibrate the galaxy names to their SIMBAD IDs, and match the names to the OSC SN hosts.
Similarly, the SFR and stellar mass $M_*$ from \cite{2016ApJ...818..182V}
as well as the age and metallicity from \cite{2002MNRAS.330..547T} are appended to the OSC SN hosts.

{For the galaxies within the coverage of SDSS, we use S{\'e}rsic half-light radius $R_{50, \rm s}$ in the NASA-SDSS Atlas catalog (NSA)\footnote{http://www.nsatlas.org} when available, which photometered 640,000 SDSS galaxies
within $z<0.15$. 
For large galaxies, e.g. $R_{50} \sim 1'-1.5^{\rm o}$, \cite{2003AJ....125..525J}
estimated the radii of the largest 656 galaxies with the 2MASS images, 
which also include the galaxies out of the coverage of SDSS.
We use the J band $R_{\rm e}$ in \cite{2003AJ....125..525J} for large galaxies instead. 
In addition, some $R_{50}$ of SLSN hosts are provided in \cite{2015ApJ...804...90L}.
Otherwise, we use the 
the half-light radius within the SDSS catalog\footnote{http://www.sdss3.org/} 
by matching the SN host galaxies with SDSS galaxies within 3''.
In the SDSS catalog, the half-light radius is estimated by the $r$-band petrosian half-light radius $R_{50, \rm petro}$ and the Petrosian 90\% radius $R_{50, \rm petro}$ following $R_{50}=R_{50, \rm petro}/(1-8\times 10^{-6}(R_{90, \rm petro}/R_{50, \rm petro})^{8.47})$
\citep{2005AJ....130.1535G}.
%However, the SDSS standard pipeline over-subtracts the sky background for large galaxies, and only gives petrosian, exponential, and de Vaucouleurs half-light radii \citep{2011AJ....142...31B}, which are more or less different from that estimated using the S{\'e}rsic profile.
%The background subtraction is improved by \cite{2011AJ....142...31B}.
%With the improved sky-subtraction techniques, the NASA-SDSS Atlas catalog(NSA) \footnote{http://www.nsatlas.org} re-photometered 640,000 SDSS galaxies within $z<0.15$, which also includes S{\'e}rsic half-light radius $R_{50, \rm s}$. 
%We thus use NSA $R_{50, \rm s}$ as our $R_{50}$ when available. 
}

We also update the offset parameters from the literatures.
\cite{2015ApJ...804...90L} calculated the projected $R_{\rm off}$ from HST images for SLSNe. 
\cite{2016A&A...589A.110A} reported the projected $R_{\rm off}$ from SN II to the nearest H II regions.
\cite{2012ApJ...759..107K} provided the deprojected $r_{\rm off}=R_{\rm off}/R_{50}$ for SN II and SN Ibc. Although projected offsets and deprojected offsets are statistically consistent, they are different for specific objects. Since most offsets in the literature and catalogs are projected, we use these deprojected offsets only when projected offsets are not available. We also convert them to $r_{\rm off}$ when $R_{50}$ is available in our catalog.

{The sample sizes for SLSNe, SN Ibc, SN  II, and SN Ia with host galaxy properties are 107, 900, 4251 and 7221, respectively. To have a consistent $z$ range as the FRB sample, we screen the samples with $z<1$. This gives the sample sizes 95, 874, 2829, and 7003 for the four types, repectively. For SN Ibc, SN II, and SN Ia, most of them have $z<1$. The excluded objects are mostly those without redshift information.}

\section{Multivariate Comparison}\label{sec:comparison}

We would like to perform a comparative study of the host galaxy properties between FRBs and other stellar transients. Since multiple parameters are involved, multivariate analysis methods are needed.
We perform two tests. First, taking all the FRBs as a whole sample, we compare it with other samples using the multivariate KS test. 
Second, for individual FRBs, we also compare them with other samples to see which type it most likely belongs to.
For easy understanding, we use the Naive Bayes method to test individual FRBs in the FRB sample to see how they may be consistent with various types of stellar transients. 
We try to classify LGRBs, SLSNe, SN Ib/Ic, SN II, SN Ia, and SGRBs with their host galaxy properties using the Naive Bayes method and then apply the same method to each FRB. 
Because galaxies evolve with redshift and the FRBs in our sample all have redshift $z<1$, {to enable a direct comparison}, in the following we compare the FRBs with the transients with redshifts $z<1$ only.

\subsection{Multivariate KS test}

\subsubsection{Method}

The classical Kolmogorov-Smirnov (KS) test compares the cumulative distribution function (CDF) of two distributions $P(<x)$ and $P(<x')$,
which could be one data sample and one model sample, or two data samples.
The largest distance between the two CDFs is defined as $D_{\rm KS}={\rm max} (P(<x)-P(<x'))$, representing the difference between the two distributions $x$ and $x'$.
The distribution of $D_{\rm KS}$ is free from the distribution of $x$ and $x'$ and independent of the direction of data ordering, i.e., the $D_{\rm KS}$ calculated from $P(<x)$ is the same as that calculated from $P(>x)$. This method is widely applied in defining the goodness of a fit or comparing two samples.

The key problem in generalizing the classical KS test to multi-dimensional
is the direction of the data ordering. 
\cite{1983MNRAS.202..615P} (P83) suggested to use the maximum absolute difference between the two samples $D_{\rm DKS}$ when all possible directions along the axes are considered.
For a two dimensional problem, the difference $D_{\rm DKS}$ values are calculated for 4 quadrants of $n^2$ origins,
$$
\left\{
\begin{array}{lr}
(x<X_i,y<Y_j) \\ 
(x<X_i,y>Y_j) \\
(x>X_i,y<Y_j) \\
(x>X_i,y>Y_j)
\end{array}
\right., ~~~~    (i,j=1,...,n)
$$
for all possible $i$ and $j$ values.  
Here $n$ is the sample size, and the method applies to comparing a data sample with a model.
The $D_{\rm DKS}$ is confirmed to be efficient for correlated samples.
For $D$ dimensions, the number of quadrants to be calculated would be $D^2n^D$, which is computationally expensive for dimensions larger than 2 and/or $n>100$.

\cite{1987MNRAS.225..155F} (FF87) proposed to use a simpler and faster method. For a two dimensional problem, the difference $D_{\rm DKS}$ is calculated for 4 quadrants of $n$ origins, 
$$
\left\{
\begin{array}{lr}
(x<X_i,y<Y_i) \\ 
(x<X_i,y>Y_i) \\
(x>X_i,y<Y_i) \\
(x>X_i,y>Y_i)
\end{array}
\right., ~~~(i=1,...,n).
$$
This method is proved not to compromise the power of the test.
With this method, only $D^2n$ quadrants should be considered for a D-dimensional size $n$ comparison, which is much lmore efficient than the P83 method.

Moreover, FF87 generalized the method to two-sample multi-dimensional KS tests by proposing to use the average $\bar{D}_{\rm DKS}$ of the two $D_{\rm DKS}$ estimated according to the data sample 1 and sample 2. 
We use the FF87 two-sample method in our analysis. 
However, the two-sample test in FF87 requires that the correlations among parameters are similar to each other in the two samples in order to use the probability $P$ distribution presented in their paper. This may not be the case in our problem. 
We thus estimate the null probability $P$ with Monte Carlo simulations.
Since our FRB sample is much smaller than our stellar transient samples, in each trial, we randomly extract the transient samples to have the same number of events as FRBs and calculate $D_{\rm mcDKS}$. We calculate the $D_{\rm mcDKS}$ 1000 times and obtain the distribution of $D_{\rm mcDKS}$. The null probability $P_{\rm DKS}$ between the FRB host galaxy sample and the transient host galaxy samples is estimated by interpolating $D_{\rm DKS}$ in the simulated $D_{\rm mcDKS}$ distribution.
{Notice that such $P_{\rm DKS}$ values from this MC simulation may have large uncertainties when both the FRB sample and the transient sample are small.}

\subsubsection{Results for galactic-scale parameters: log $M_*$ and log sSFR}

\begin{figure}[htb]
\centering
\includegraphics[scale=0.6]{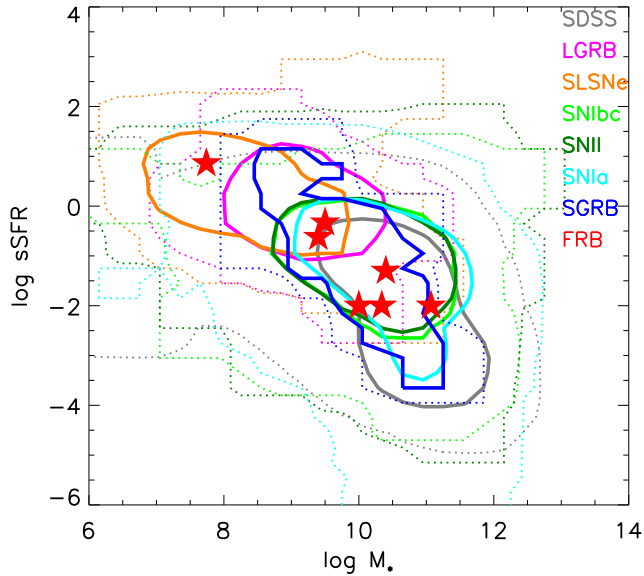}
\caption{Comparison of FRB hosts(stars) with other stellar transients, as well as SDSS galaxies.
The solid lines and dotted lines represent 1 $\sigma$ and 3 $\sigma$ region of each stellar transients.}
\label{fig:compare}
\end{figure}

The most common parameters available for transient host galaxies are stellar mass log $M_*$ and log sSFR, regardless of whether the transients are well located in the sub-galactic scale. Thus, log $M_*$ and log sSFR present the largest samples, both for FRBs and other stellar transients. {Seven} FRB host galaxies have the information for log $M_*$ and log sSFR. 
The numbers for each type transient with both log $M_*$ and log sSFR are listed in the second column of Table \ref{ndKS}.
{The FRBs (stars) are compared with various stellar transients in the log $M_*$ -- log sSFR plane in Fig. \ref{fig:compare}.
The stellar transient hosts and SDSS galaxies are presented as solid and dotted contours, 
with 1$\sigma$ (68\%) and 3$\sigma$ (99.7\%) confidence levels, respectively. 
The 1 $\sigma$ contours seem to occupy three regions in the log sSFR$-$log $M_*$ diagram. 
1. {$7<{\rm log} M_*<10.5$ \& $-1<{\rm log~sSFR}<1$}: LGRB (magenta) and SLSNe (orange) hosts occupy the strongest star formation region, while SLSN hosts have smaller stellar mass log $M_*$ than LGRB hosts.
2. $8.5<{\rm log} M_*<11.5$ \& $-3<{\rm log~sSFR}<0$: SN Ibc (green) and SN II (dark green) hosts reside in the middle region in the log sSFR -- log $M_*$ plane. This region also consists of the star-forming galaxy of SDSS and SN Ia (cyan) hosts. 
3. $10<{\rm log} M_*<12$ \& ${\rm log~sSFR}<-2$: There are also passive galaxy components in SN Ia hosts and SDSS (grey) galaxies. 
SGRB (blue) hosts occupy all three regions in the log sSFR$-$log $M_*$ diagram.
For FRBs, the FRB 121102 host is located in the region with the least stellar mass and strongest star formation.
The FRB 181112, FRB 190102 and FRB 190608 hosts reside in the middle region. 
The hosts of FRB 180924, FRB 190523, and FRB 180916.J0158+65 are in the joint region between the star formation galaxies and passive galaxy component of SDSS galaxies, SN Ia hosts, and SGRB hosts.
{Except FRB 121102, all other FRB hosts are consistent with the hosts of SNe Ibc, SNe II, SNe Ia and SGRBs within the 68\% region.}
}

We make a two-dimensional KS test between FRBs and other transients
in the log $M_*$ - log sSFR space. 
The $D_{\rm 2KS}$ and the null probability values are presented in the third and fourth column of Table \ref{ndKS}. It turns out that the whole FRB sample rejects the origin similar to LGRBs and SLSNe with a significant level of {0.02}, while other origins similar to SN Ibc, SN II, SN Ia, and SGRBs are still consistent with the FRB sample.
{Although the SGRB sample is small and the relation between its $D_{\rm DKS}$ and $P_{\rm DKS}$ has significant uncertainties, the $D_{\rm DKS}$ of SGRBs is the smallest among all the transients, indicating similarity or more consistency of FRBs with SGRBs.}
 
We also examine the differences among different types of transients in the log $M_*$ - log SFR space (i.e. sSFR is replaced by SFR). The results are similar: that the FRB hosts and LGRB/SLSN hosts are significantly different, suggesting that the FRB host galaxy as a whole disfavors the LGRB and SLSNe origin.
Similarly, the FRB hosts are also consistent with those of other transients in the log $M_*$ - log SFR space.

{We note that different SFR indicators represents stars with different age \citep{2012ARA&A..50..531K}, which introduce uncertainties to the SFR and sSFR in our sample. Moreover, SFR indicators, including emission lines, UV and IR emission, are usually influenced by the emission from AGNs. Many FRBs, including FRB 190102, FRB 180924, and FRB 190608, are located in the galaxies with AGN or LINER emissions \citep{bhandari2020}. In these cases, the log SFR and log sSFR values should be considered as the upper limits. As a result, to compare their environments more accurately, sub-galactic parameters should be taken into account.}

\begin{table}[!bht] \scriptsize
\begin{center}
\caption{Multivariate KS test results}
\label{ndKS}
\begin{tabular}{l|ccc|ccc|ccc}
\hline
 & \multicolumn{3}{c|}{$M_*$ \& sSFR}  & \multicolumn{3}{c|}{$M_*$, sSFR, $R_{\rm off}$} & \multicolumn{3}{c}{$M_*$,sSFR} \\
 & \multicolumn{3}{c|}{}  & \multicolumn{3}{c|}{} & \multicolumn{3}{c}{$R_{50}$,$R_{\rm off}$,$F_{\rm light}$} \\
 \hline
name & No. & $D_{\rm DKS}$ & $P_{\rm DKS}$  & No. & $D_{\rm DKS}$ & $P_{\rm DKS}$ & No. & $D_{\rm DKS}$ & $P_{\rm DKS}$  \\
FRB & 7 & & & 6 & & & 2 & & \\\hline 
LGRB & 49 & 0.62 & 0.02 & 38 & 0.66 & 0.03 & 17 & 0.49 & 0.99\\
SLSNe & 81 & 0.72 & 0.002 & 16 & 0.76 & 0.01 & 9 & 0.78 & 0.28\\
SNIbc & 298 & 0.38 & 0.68 & 250 & 0.41 & 0.85 & 55 & 0.70 & 0.95\\
SNII & 1058 & 0.43 & 0.56 & 919 & 0.46 & 0.79 & 98 & 0.76 & 0.92\\
SNIa & 2286 & 0.46 & 0.42 & 1815 & 0.52 & 0.53 & 74 & 0.93 & 0.30\\
SGRB & 18 & 0.29 & 0.81 & 16 & 0.36 & 0.80 & 11 & 0.64 & 0.76\\
\hline
\end{tabular}
\end{center}
\end{table}

\subsubsection{Results for combined galactic and sub-galactic parameters: log $M_*$, log sSFR, log $R_{50}$, log $R_{\rm off}$, and $F_{\rm light}$}

The positions of the transients within their host galaxies also provide important {and usually more accurate} information about the origin of the transients. {The offset is available for all the nine FRBs. However, the positional uncertainty of FRB 181112 is larger than the size of the galaxy. We do not include it in our analysis here. Although the positional uncertainty of FRB 190523 is also large, its offset is even more significant. We thus keep it in the sample. There are six FRBs with log $M_*$, log sSFR and log $R_{\rm off}$ in our analysis. We make a three-dimensional KS test between these six FRBs and other transients in the log $M_*$ $-$log sSFR $-$ log $R_{\rm off}$ space. The results are presented in the 5-7 columns of Table \ref{ndKS}. The conclusion is similar to what we have in the previous section: The FRB hosts as a whole disfavor the LGRB and SNSL origin.} 

{On the other hand, the offset is also controlled by the size $R_{50}$ and the gravitational well of the host. And, due to the irregularity of some hosts, such as the host of FRB 121102, $F_{\rm light}$ better represents the local environments of the transients\cite{2006Natur.441..463F, 2012MNRAS.424.1372A}. A more complete comparison of galactic and sub-galactic environments should include five parameters:  log $M_*$, log sSFR, log $R_{50}$, log $R_{\rm off}$, and $F_{\rm light}$. }
The numbers of each type of transient with all the five parameters are listed in the eighth column of Table \ref{ndKS}.
And the $D_{\rm 2KS}$ values and the null probability values of the five-dimensional KS test between FRBs and other transients are presented in the ninth and tenth columns of Table \ref{ndKS}. However, due to the small size, the FRB sample does not show a significant difference with respect to any other type of stellar transient. 
{A larger sample and/or a more detailed analysis of the FRB hosts may help to narrow down the preferred transients if any, or suggest that FRBs are indeed consistent with a broad range of transient types. }

\subsection{Naive Bayes}

\begin{figure*}[htb]
\centering
\includegraphics[width=0.45\textwidth]{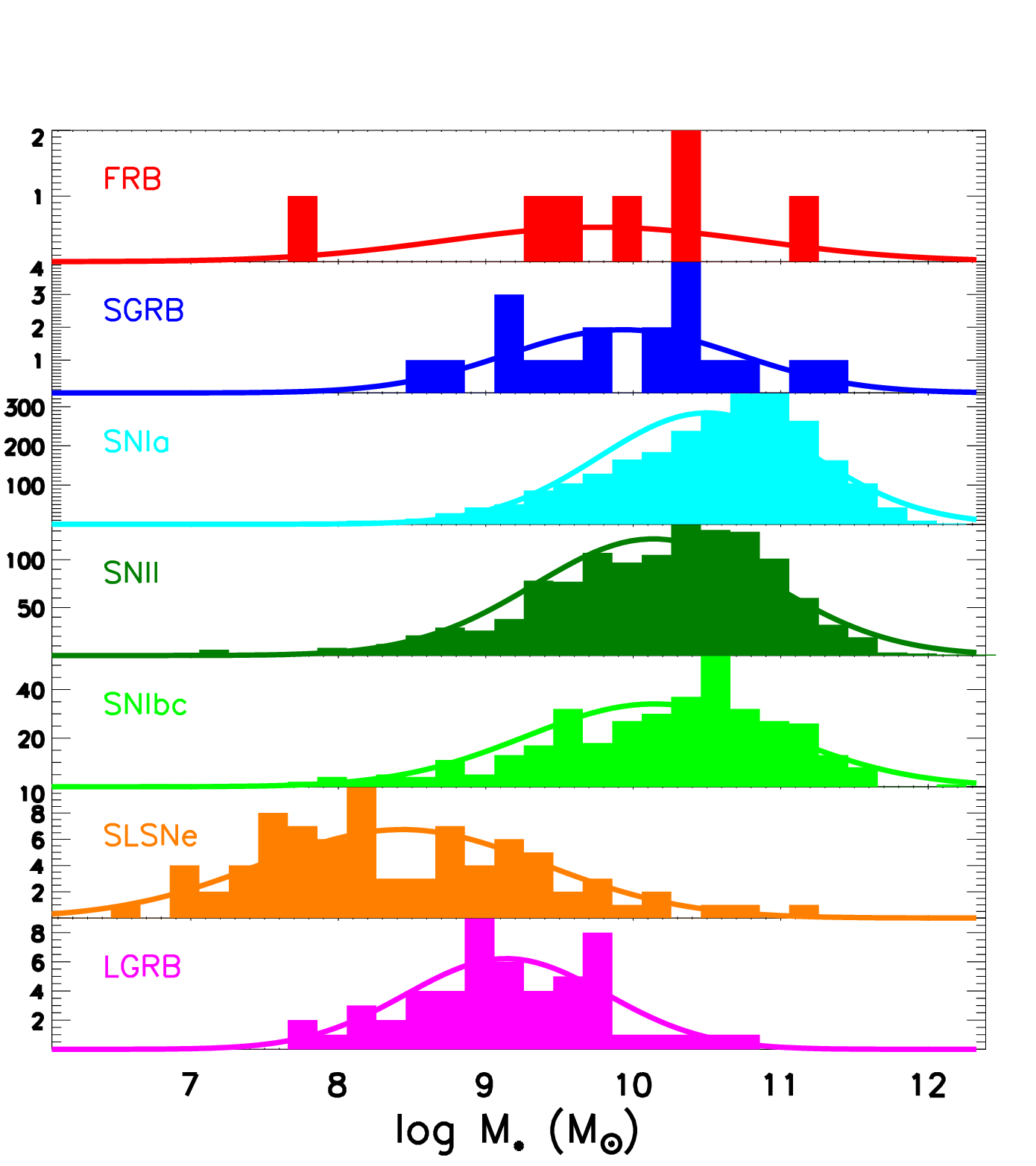}
\includegraphics[width=0.45\textwidth]{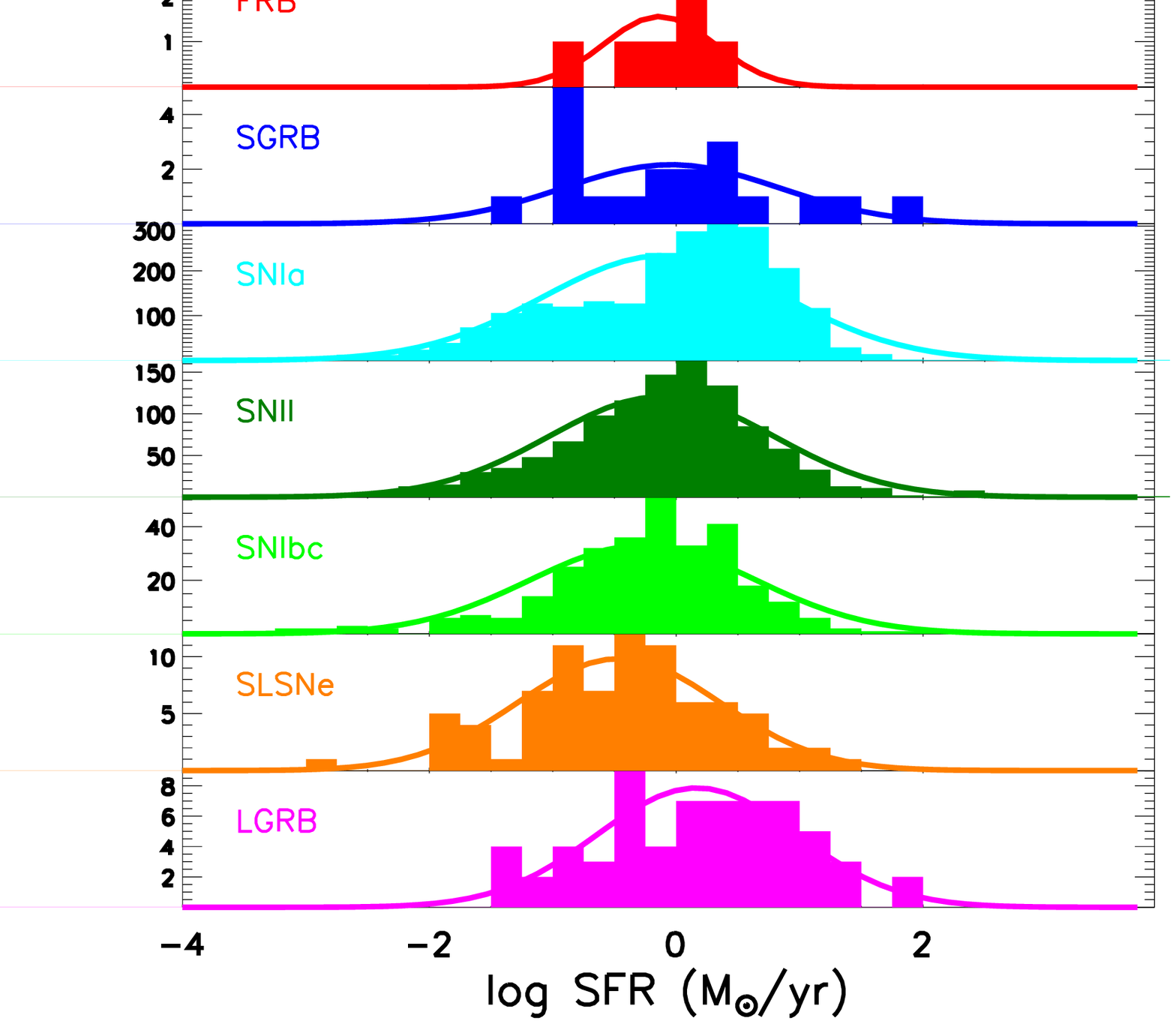}
\includegraphics[width=0.45\textwidth]{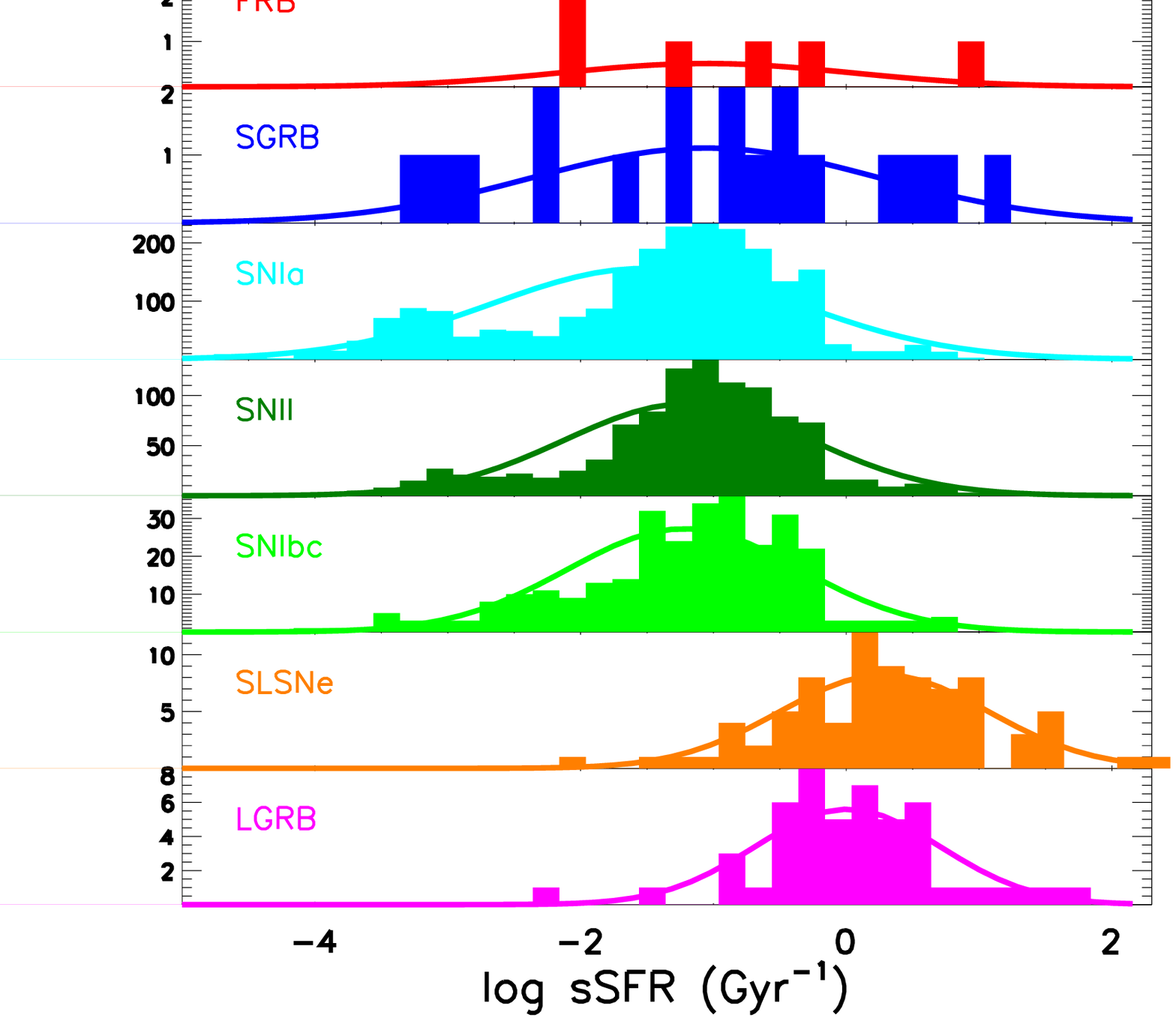}
\includegraphics[width=0.45\textwidth]{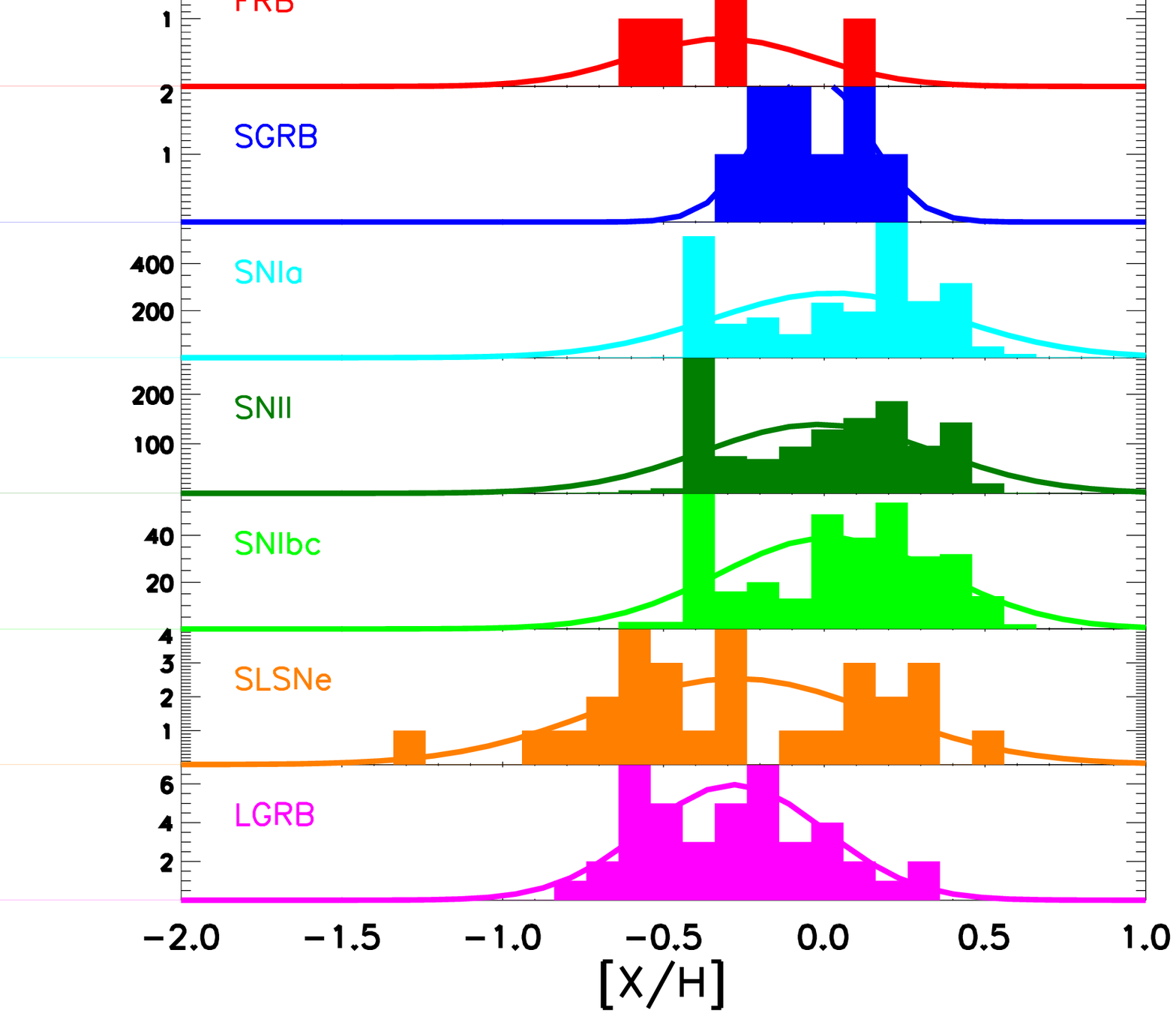}
\caption{The distributions and fitting results of LGRBs (magenta), SLSNe (orange), SN Ibc (green), SN II(dark green), SN Ia (cyan), SGRBs (blue), and FRB(red). }
\label{fig:NBfit}
\end{figure*}

\begin{figure*}[htb]
\centering
\includegraphics[width=0.45\textwidth]{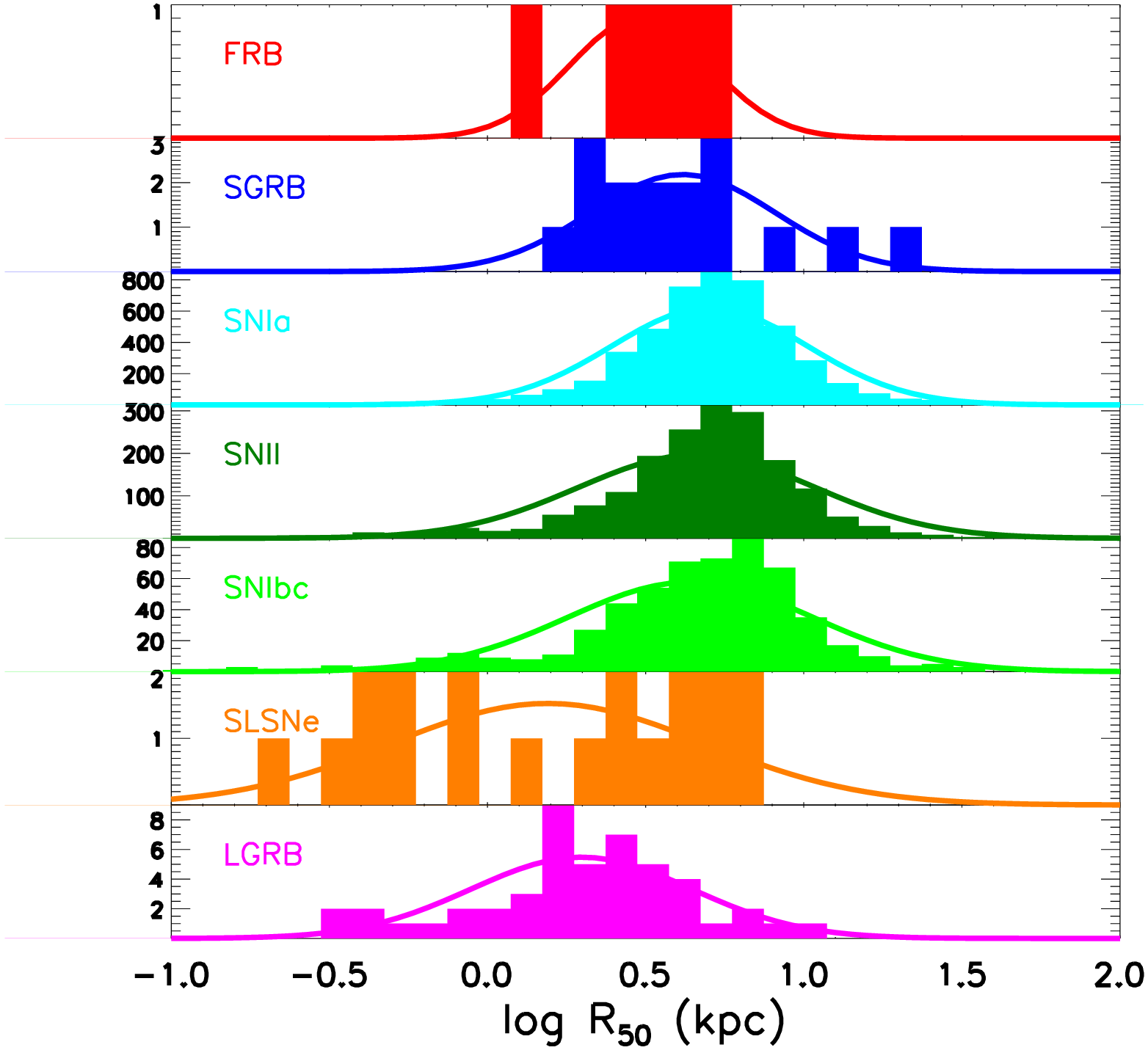}
\includegraphics[width=0.45\textwidth]{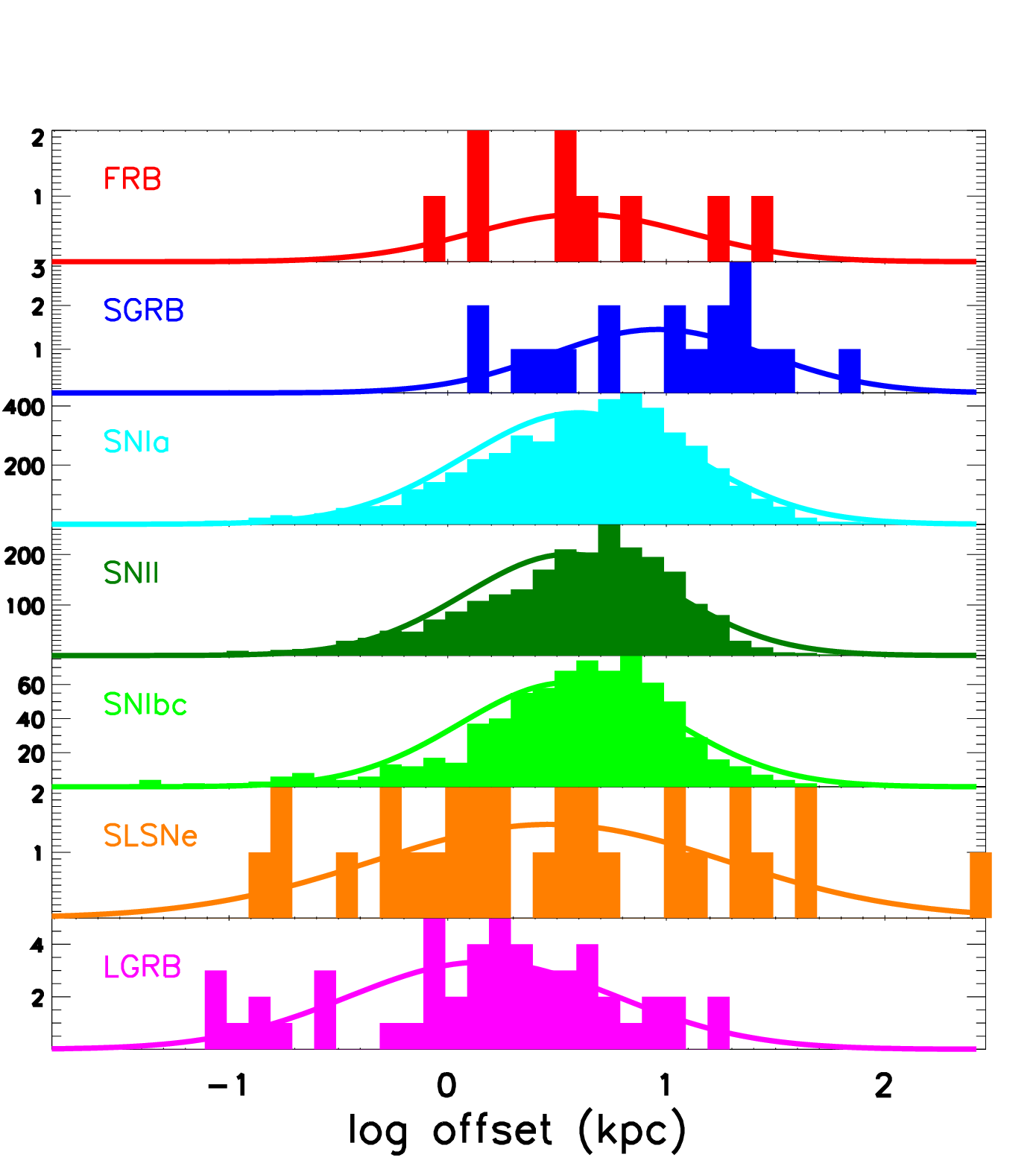}
\includegraphics[width=0.45\textwidth]{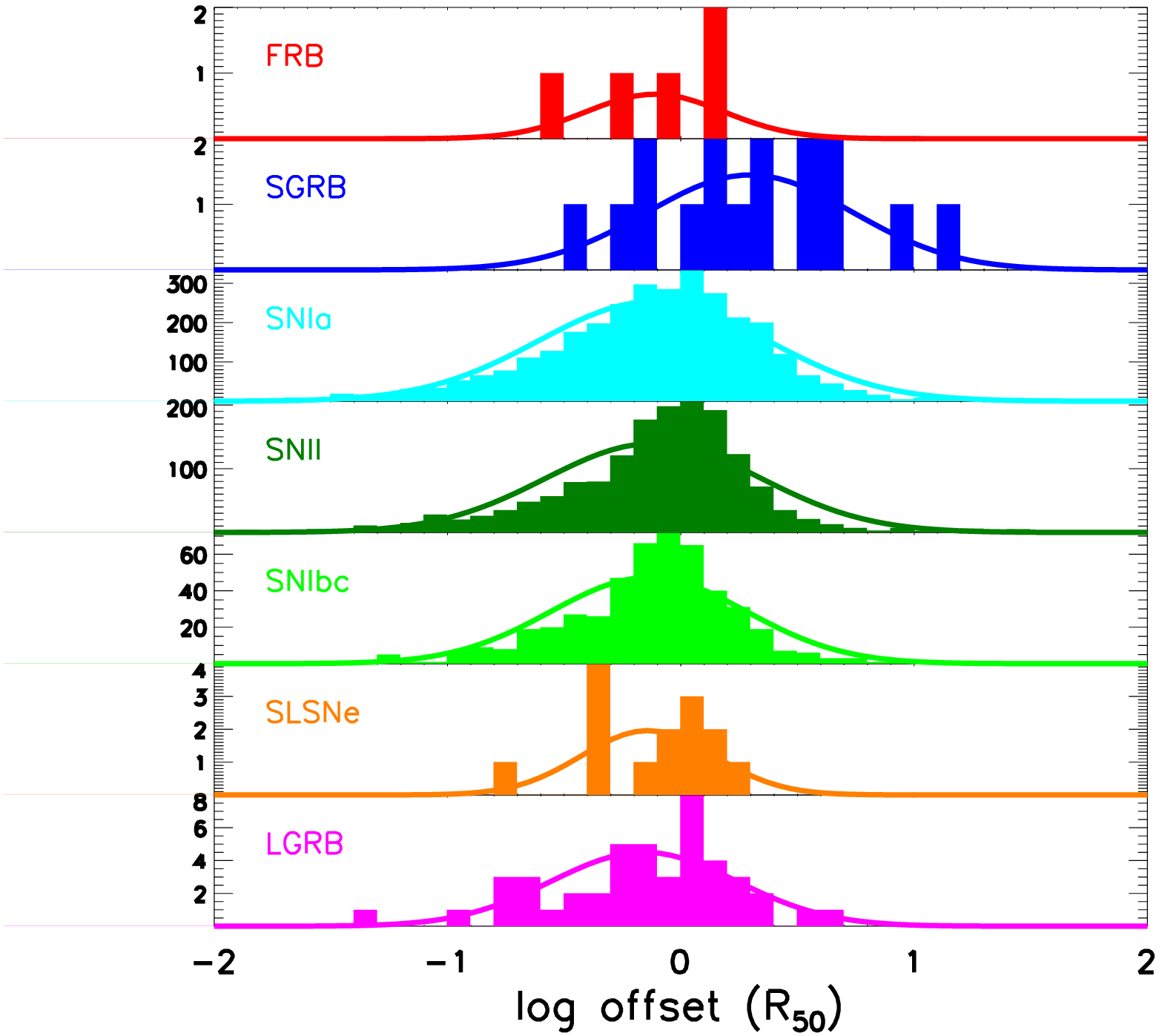}
\includegraphics[width=0.45\textwidth]{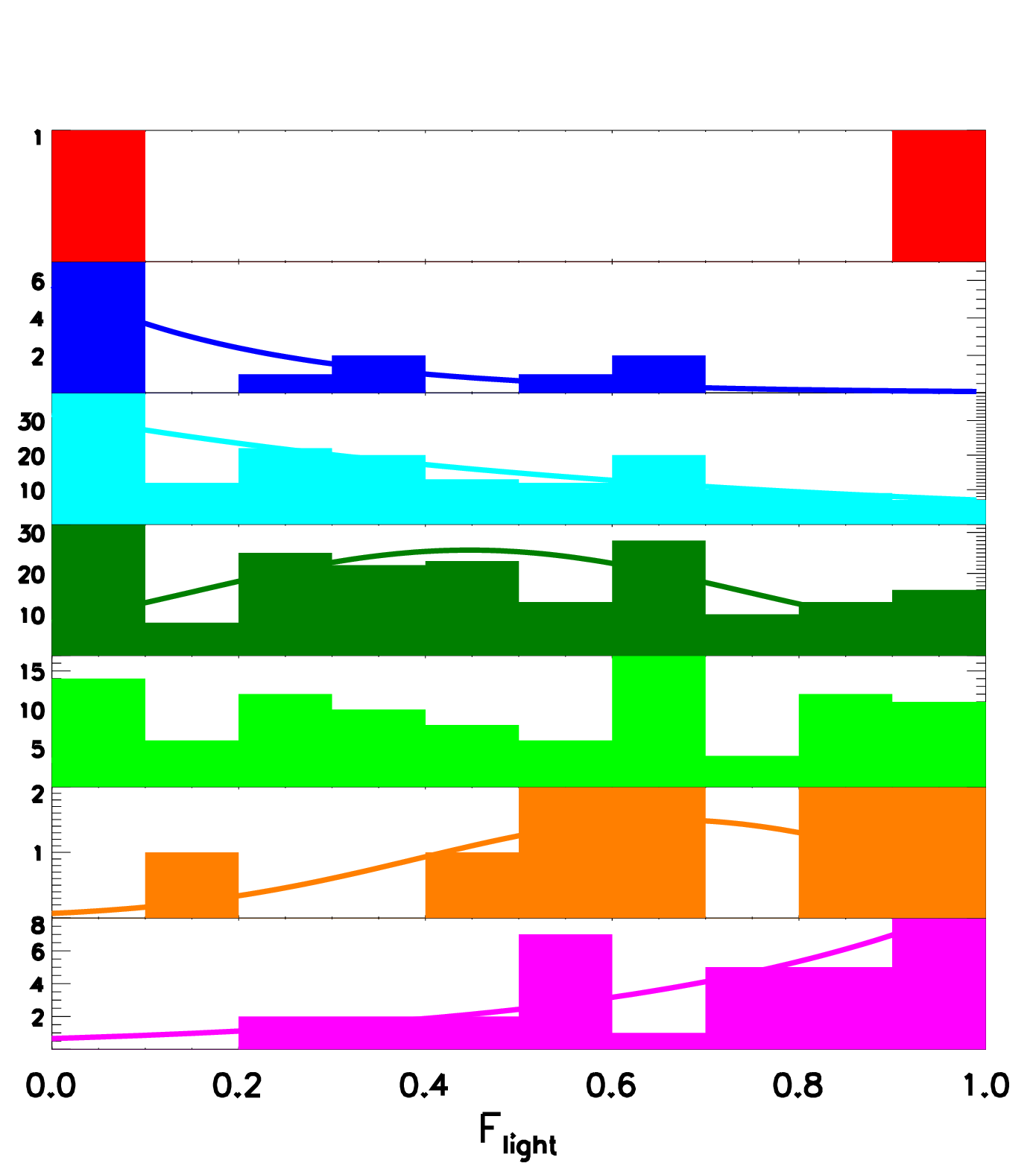}
\center{Figure \ref{fig:NBfit} -- continued. }
\end{figure*}

\begin{table*}[tb] \scriptsize
\begin{center}
\caption{Fitting results of each parameter for All parameters. }
\label{tbfit}
\begin{tabular}{l|ccccccccc}
\hline
  & \multicolumn{2}{c}{log SFR (M$_{\odot}$ yr$^{-1}$)} & \multicolumn{2}{c}{log sSFR (Gyr$^{-1}$)} & \multicolumn{2}{c}{log $M_*$ ($M_{\odot}$)} & \multicolumn{2}{c}{$\rm [X/H]$}\\
 & $\mu \pm \sigma$ & $D_{\rm KS}/P_{\rm KS}$ & $\mu \pm \sigma$ & $D_{\rm KS}/P_{\rm KS}$ & $\mu \pm \sigma$ & $D_{\rm KS}/P_{\rm KS}$ & $\mu \pm \sigma$ & $D_{\rm KS}/P_{\rm KS}$\\\hline
LGRB & $0.17 \pm 0.81$  & 0.09/0.69 & $0.01 \pm 0.70$  & 0.10/0.63 & $9.14 \pm 0.68$  & 0.06/0.99 & $-0.28 \pm 0.28$  & 0.10/0.78 \\
SLSNe & $-0.46 \pm 0.82$  & 0.05/0.98 & $0.29 \pm 0.78$  & 0.07/0.86 & $8.43 \pm 0.96$  & 0.10/0.34 & $-0.27 \pm 0.44$  & 0.14/0.61 \\
SNIbc & $-0.27 \pm 0.92$  & 0.08/0.03 & $-1.22 \pm 0.88$  & 0.08/0.05 & $10.14 \pm 0.87$  & 0.09/0.01 & $0.02 \pm 0.34$  & 0.11/0.00 \\
SNII & $-0.12 \pm 0.91$  & 0.06/0.00 & $-1.19 \pm 0.92$  & 0.10/0.00 & $10.14 \pm 0.81$  & 0.07/0.00 & $-0.02 \pm 0.36$  & 0.12/0.00 \\
SNIa & $-0.10 \pm 0.98$  & 0.10/0.00 & $-1.52 \pm 1.16$  & 0.13/0.00 & $10.50 \pm 0.73$  & 0.07/0.00 & $0.03 \pm 0.37$  & 0.12/0.00 \\
SGRB & $-0.04 \pm 0.88$  & 0.11/0.96 & $-1.06 \pm 1.30$  & 0.10/0.99 & $9.93 \pm 0.79$  & 0.11/0.96 & $-0.03 \pm 0.16$  & 0.19/0.87 \\
FRB & $-0.14 \pm 0.45$  & 0.26/0.68 & $-1.06 \pm 1.09$  & 0.24/0.77 & $9.78 \pm 1.06$  & 0.22/0.85 & $-0.31 \pm 0.28$  & 0.21/0.96 \\
\hline
  & \multicolumn{2}{c}{log $R_{50}$ (kpc)} & \multicolumn{2}{c}{log offset (kpc)} & \multicolumn{2}{c}{log offset ($R_{50}$)} & \multicolumn{2}{c}{$F_{\rm light}$} \\
 & $\mu \pm \sigma$ & $D_{\rm KS}/P_{\rm KS}$ & $\mu \pm \sigma$ & $D_{\rm KS}/P_{\rm KS}$ & $\mu \pm \sigma$ & $D_{\rm KS}/P_{\rm KS}$ & $\mu \pm \sigma$ & $D_{\rm KS}/P_{\rm KS}$\\\hline
LGRB & $0.30 \pm 0.35$  & 0.11/0.62 & $0.16 \pm 0.61$  & 0.11/0.60 & $-0.16 \pm 0.40$  & 0.09/0.83 & $2.62^*$  & 0.11/0.82 \\
SLSNe & $0.19 \pm 0.50$  & 0.17/0.60 & $0.47 \pm 0.81$  & 0.11/0.84 & $-0.14 \pm 0.28$  & 0.14/0.92 & $0.66 \pm 0.27$  & 0.13/0.99 \\
SNIbc & $0.64 \pm 0.39$  & 0.10/0.00 & $0.55 \pm 0.49$  & 0.08/0.00 & $-0.14 \pm 0.42$  & 0.09/0.00 & $0.39 \pm 1.09$  & 0.35/0.00 \\
SNII & $0.66 \pm 0.38$  & 0.11/0.00 & $0.56 \pm 0.48$  & 0.07/0.00 & $-0.12 \pm 0.46$  & 0.09/0.00 & $0.45 \pm 0.30$  & 0.06/0.39 \\
SNIa & $0.70 \pm 0.31$  & 0.07/0.00 & $0.60 \pm 0.52$  & 0.05/0.00 & $-0.12 \pm 0.49$  & 0.05/0.00 & $-1.53^*$  & 0.16/0.00 \\
SGRB & $0.62 \pm 0.29$  & 0.19/0.60 & $0.96 \pm 0.49$  & 0.16/0.70 & $0.30 \pm 0.44$  & 0.08/1.00 & $-4.35^*$  & 0.46/0.00 \\
FRB & $0.48 \pm 0.22$  & 0.24/0.91 & $0.61 \pm 0.50$  & 0.14/0.99 & $-0.11 \pm 0.29$  & 0.20/0.97 & $-$  & $-$ \\
\hline
\end{tabular}
\end{center}
\footnotesize
$^*$ For $F_{\rm light}$, this is the index $\gamma$ of the exponential distribution.
\end{table*}

It is possible that there might exist different sub-types of FRBs with distinct origins, so that different FRBs may fall into different distributions. 
While the multivariate KS test compares FRBs with other stellar transients as a whole sample, it cannot test how individual FRBs fall into the distribution of a certain type of transient. 
We adopt the Naive Bayes method to perform such a task.

\subsubsection{Method}
The Naive Bayes method is a classification method based on the Bayes theorem and the assumption that parameters are not correlated, i.e.
\begin{eqnarray}
P({\rm T}|\{x\}) = \frac{P(\{x\}|{\rm T})P({\rm T})}{P(\{x\})}, \\
P(\{x\}|{\rm T})= \prod\limits_{i}P(x_i|{\rm T}),
\label{nb}
\end{eqnarray}
where $P({\rm T}|\{x\})$ is the posterior probability for one object with parameter set $\{x\}$ 
to have a type T, which could be LGRBs, SLSNe, SN Ibc, SN II, SN Ia, or SGRBs;
$P(\{x\}|\rm T)$ is the likelihood for one object with a type T to have a parameter set $\{x\}$, $P(\rm T)$ is the prior probability of one object to be in type T, and $P(\{x\})$ is the probability for one object to have a parameter set $\{x\}$.
Although the assumption that parameters are uncorrelated is ``naive'', 
the results are amazingly good \citep{NBhand2001, NBbroos2011}.

The implementation of Naive Bayes in our problem follows the following steps: 
\begin{enumerate}{\itemindent=3em}
    \item Estimate the likelihood $P(x_{\rm i}|{\rm T})$ for each parameter $x_{\rm i}$ and type T with the known stellar transient samples;
    \item Estimate the prior $P({\rm T})$, usually by the size of the sample. However, for FRBs, we do not have any prior information about the preference to specific types. We thus use equal prior for all types of transients.
    \item Calculate the posterior probability following equation (\ref{nb}) for each type of transient T;
    \item Normalize the posterior probability $P({\rm T}|\{x\})$ by requiring $\sum\limits_{j} P(T_{\rm j}|\{x\})=1$ for each FRB, where $T_{\rm j}$ stands for different transient types;
    \item The type of the FRB is assigned to the one with the highest posterior probability.
\end{enumerate}

We estimate the likelihood for each parameter $x_i$ and type T, $P(x_i|{\rm T})$, with the observed sample in Section \ref{chp_sample}. Gaussian distributions are assumed for most parameters, while exponential distributions $P(x) \propto {\rm exp}(-x)$ are assumed for $F_{\rm light}$ for LGRB, SN Ia and SGRB.  
The fitting results are presented in Table \ref{tbfit} and Fig. \ref{fig:NBfit}. 
{The distribution of FRB is also presented as red for comparison}.

\subsubsection{Results}

\begin{figure}[htb]
\centering
\includegraphics[width=0.5\textwidth]{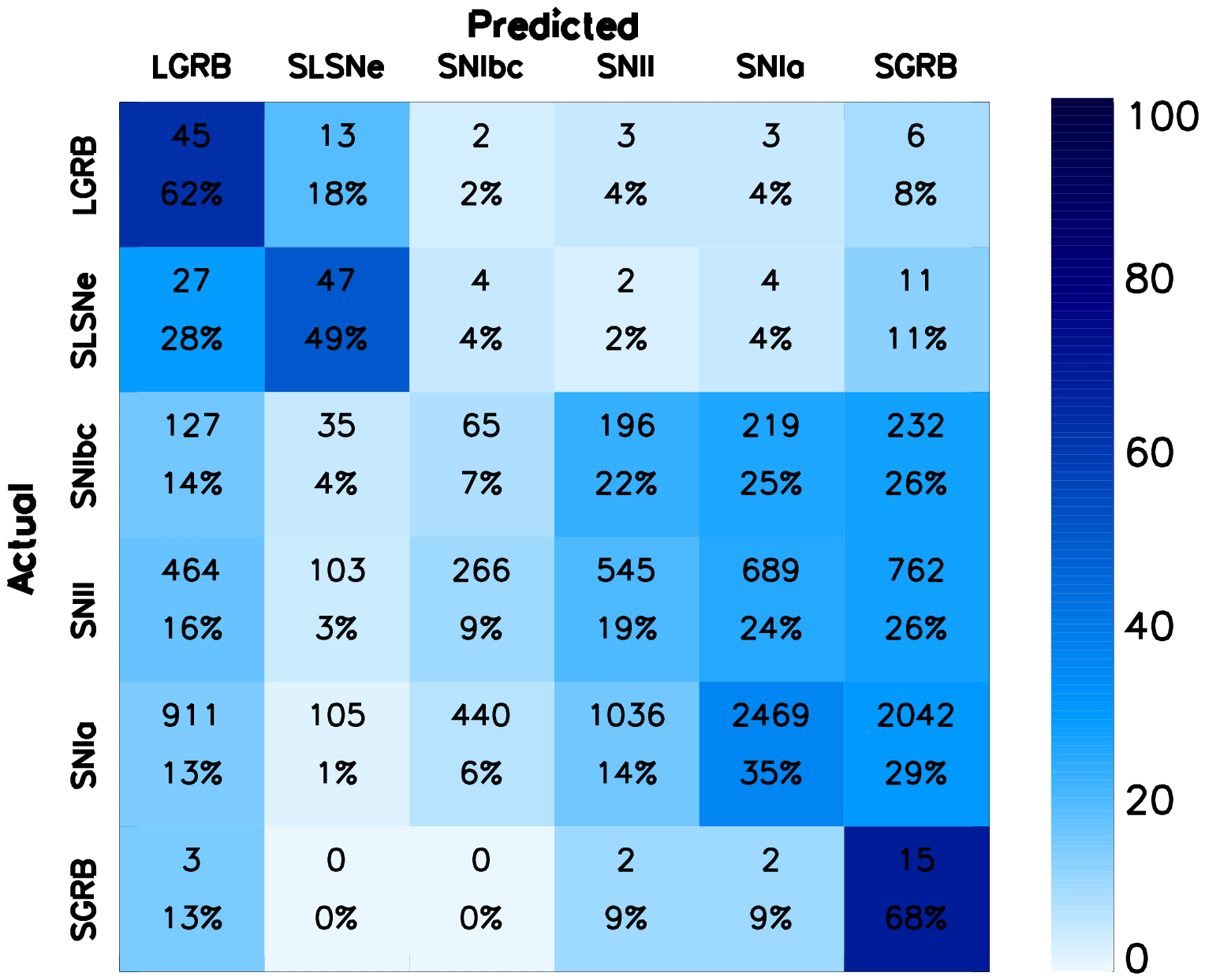}
\caption{Confusion matrix of Naive Bayes methods.}
\label{figcmatrix}
\end{figure}

The Naive Bayes results for the stellar transients are presented in Table \ref{figcmatrix}.
It shows the number of known LGRBs, SLSNe, SN Ibc, SN II, SN Ia and SGRBs classified as each type with Naive Bayes (NB).
It turns out that most of LGRBs, SLSNe, and SGRBs can be identified with their host galaxy properties. 
However, about half of SN Ibc, SN II, SN Ia may be mis-classified as SGRBs, suggesting that the host galaxy properties of these transients are not that different from those of SGRBs.

\begin{figure}[htb]
\centering
\includegraphics[width=0.5\textwidth]{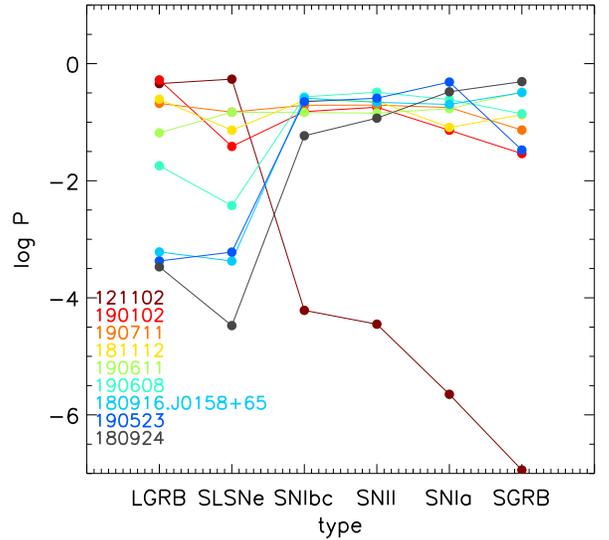}
\caption{The probability of FRB as various stellar transients.}
\label{fig:ptype}
\end{figure}

We apply the same method to individual FRBs. The results are presented in Table \ref{tbpfrb} and Fig. \ref{fig:ptype}. Following conclusions can be drawn:
The host of FRB 121102 has 90\% probability to belong to the SLSN sample and 10\% probability to belong to the LGRB sample, and its probabilities to belong to SN Ibc, SN II, SN Ia, and SGRB samples are significantly smaller. 
FRB 180916.J0158+65 has a very low probability to belong to the LGRB or SLSN samples, but is consistent with being belong to either of the  
SN Ibc, SN II, SN Ia or SGRB samples.
FRB 180924 also has a very low probability to belong to the LGRB/SLSN sample but has a fairly high probability belonging to SN Ia or SGRB samples.
FRB 181112 has a reasonable probability to belong to any sample, due to its mild SFR and log $M_*$ and the lack of sub-galactic information.
FRB 190523 also disfavors a LGRB/SLSN origin but is consistent with the SN Ibc, SN II or SN Ia origins. Its consistency with the SGRB sample is marginal.

\begin{table}[!htb] \scriptsize
\begin{center}
\caption{Probability of FRBs as stellar transients}
\label{tbpfrb}
\begin{tabular}{l|cccccc}
\hline
FRB name  & LGRB & SLSNe & SNIbc & SNII & SNIa & SGRB \\
121102 & 0.46 & 0.54 & 6e-5 & 4e-5 & 2e-6 & 1e-7\\
180916.J0158+65 & 6e-4 & 4e-4 & 0.26 & 0.22 & 0.20 & 0.32\\
180924 & 3e-4 & 3e-5 & 0.06 & 0.12 & 0.33 & 0.49\\
181112 & 0.25 & 0.07 & 0.24 & 0.23 & 0.08 & 0.13\\
190102 & 0.53 & 0.04 & 0.15 & 0.18 & 0.07 & 0.03\\
190523 & 4e-4 & 6e-4 & 0.22 & 0.26 & 0.48 & 0.03\\
190608 & 0.02 & 4e-3 & 0.27 & 0.33 & 0.24 & 0.14\\
190611 & 0.07 & 0.15 & 0.15 & 0.14 & 0.17 & 0.33\\
190711 & 0.21 & 0.15 & 0.19 & 0.20 & 0.18 & 0.07\\
\hline
\end{tabular}
\end{center}
\end{table}

\section{Conclusion and Discussion}

In this paper, we compare the properties of the FRB host galaxies with those of various stellar transients, including LGRBs, SLSNe, SN Ibc, SN II, SN Ia, and SGRBs. 
Considering the FRBs with host galaxies as a whole, a multivariate KS test reveals that FRB hosts are not consistent with the hosts of LGRBs and SLSNe in the significant level of 0.02. Due to the small sample statistics, they are still consistent with the distribution of the hosts of all other transients. 
Comparing the FRBs hosts with the stellar transient hosts individually, we find that
FRB 121102 tends to have a similar origin as LGRBs and SLSNe, while
FRB 180924, FRB 190523, FRB 180916.J0158+65 and {FRB 190608} are more similar to SN Ibc, SN II, SN Ia, or SGRBs.
{FRB 190102, FRB 190711, FRB 181112 and FRB 190611} do not show obvious preference to any type.

The results in our study may shed light on the unknown energy source of FRBs. The first implication is that repeating FRBs and apparently non-repeating FRBs do not show dichotomy in terms of host galaxy properties. In fact, two active repeaters FRB 121102 \citep{spitler16,chatterjee17,marcote17,tendulkar17} and FRB 180916.J0158+65 \citep{chime19-8repeater,marcote20} show opposite properties in terms of host galaxy properties and cannot be grouped into the same type of host galaxy categories. {Some apparently non-repeating FRBs, such as FRB 190608,} share similar properties as FRB 180916, which is consistent with the speculation that most{(or at least some)} apparently non-repeating FRBs may be repeating ones \citep{lu19,ravi19b}.

The leading FRB source model invokes magnetars as the power source to produce repeating bursts. There are two versions of this model. One version invokes rapidly spinning young magnetars that are produced in extreme stellar transients such as GRBs and SLSNe. The main motivation is that the host galaxy of FRB 121102 resembles those of LGRBs and SLSNe \citep{metzger17,nicholl17,2019ApJ...879....4W}. The fact that all other FRB hosts do not resemble that of FRB 121102 disfavors the simplest version of this proposal. A possible fix of this proposal is to introduce rapidly spinning magnetars born from binary neutron star (BNS) mergers \citep{margalit19,wang20}. In order to make this scenario to work, one needs to require that the rapidly spinning magnetars made from BNS mergers should be much more abundant than those made from LGRBs and SLSNe. 
Comparing the event rate densities of BNS mergers, LGRBs and SLSNe \citep[e.g.][]{sun15,GW170817,nicholl17}, this may be possible if a significant fraction of BNS mergers leave behind stable neutron stars \citep[e.g.][]{gao16}. However, if this fraction is very low as required if GW170817 leaves behind a black hole \citep{margalit19}, the fast magnetar model may fail to explain the small fraction of LGRB/SLSN-like hosts in FRB samples. The second version of the magnetar model invokes emission (e.g. giant flares) of slowly rotating magnetars like the ones observed in the Galaxy \citep[e.g.][]{popov10,kulkarni14,katz14}. The births of these magnetars do not require extreme explosions such as GRBs and SLSNe \citep[e.g.][]{beniamini19}. If this is the case, the host galaxy distribution may be more analogous to that of SN II. {All FRBs but FRB 121102} are consistent with this scenario (Fig.\ref{fig:ptype}). In order to interpret FRB 121102, the more extreme channel of forming rapid magnetars is still needed. So we conclude that the magnetar model would work, only if both fast magnetars produced in extreme explosions and slow magnetars produced in regular channels \citep{beniamini19} can both produce FRBs. In any case, since the birth rate of these magnetars is very high \citep{beniamini19}, an additional factor is needed to select a small fraction of magnetars to produce FRBs \citep[e.g.][]{ioka20}.

Some non-magnetar models may be accommodated by the data. The cosmic comb model \citep{zhang17} invokes a variety of possible donors to reshape the magnetosphere of neutron star magnetosphere. These events can in principle occur in a variety of host galaxies with a variety of local environments. 
{However, most other models are  consistent with the FRBs except FRB 121102.}
The pre-merger BNS interaction model \citep{zhang20} would predict host galaxy type and local environments similar to those of SGRBs. 
Models invoking white dwarf mergers \citep{kashiyama13} or white dwarf accretion \citep{gu16} may have hosts similar to SN Ia. 
The $\sim 16$-day periodicity of FRB 180916.J0158+65 \cite{chime-periodicity} may require a binary systems (e.g. \cite{ioka20,lyutikov20,katz20,dai20}, see also \cite{2020arXiv200312509B}). These systems may have a host galaxy or local environment similar to intermediate stellar populations. 

The constraints on the FRB source models are limited by the small sample of FRBs with host galaxy observations. Continued localization campaigns of FRBs by ASKAP and other facilities will increase the sample of FRB hosts significantly in the upcoming years. 
On the other hand, the intrinsic degeneracy of the host galaxy properties among many stellar explosions (e.g. SN Ib/c, SN II, SGRBs) makes it difficult to identify the origin of FRBs based on the host galaxy properties alone. Additional information (e.g. multi-frequency counterparts, periodicity) is needed to eventually pin down the origin of FRBs.

\acknowledgements
{We thank the anonymous referee for helpful suggestions and comments.} 
YL thanks Dongdong Shi, Qiang Yuan and Jie Zheng for helpful discussion.
YL is supported by the KIAA-CAS Fellowship, which is jointly supported by Peking University and Chinese Academy of Sciences. Her work is also partially supported by the China Postdoctoral Science Foundation (No. 2018M631242).

Funding for SDSS-III has been provided by the Alfred P. Sloan Foundation, the Participating Institutions, the National Science Foundation, and the U.S. Department of Energy. The SDSS-III web site is http://www.sdss3.org.

SDSS-III is managed by the Astrophysical Research Consortium for the Participating Institutions of the SDSS-III Collaboration including the University of Arizona, the Brazilian Participation Group, Brookhaven National Laboratory, University of Cambridge, University of Florida, the French Participation Group, the German Participation Group, the Instituto de Astrofisica de Canarias, the Michigan State/Notre Dame/JINA Participation Group, Johns Hopkins University, Lawrence Berkeley National Laboratory, Max Planck Institute for Astrophysics, New Mexico State University, New York University, Ohio State University, Pennsylvania State University, University of Portsmouth, Princeton University, the Spanish Participation Group, University of Tokyo, University of Utah, Vanderbilt University, University of Virginia, University of Washington, and Yale University.

%\bibliographystyle{apj}
%\bibliography{references}

%\end{document}

\end{document}